# Where do ideas come from?

Abraham Loeb

Ideas are nurtured by informal dialogues in environments where mistakes are tolerated and critical thinking is encouraged.

———

June 26, 2018

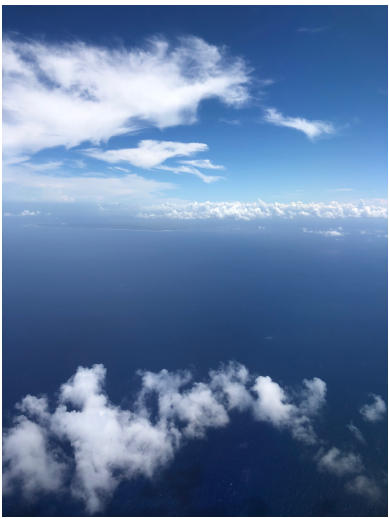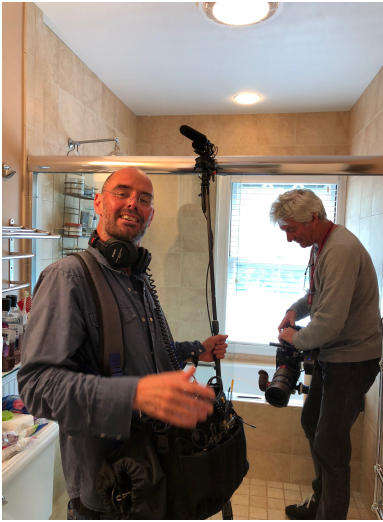

*Photos credit:* A. Loeb (2018)

Recently a Dutch TV crew came to my home for an interview about my latest research in Astronomy. When I told them that I get many of my new ideas in the shower, they decided to film a scene showing the shower still running and me rushing from the bathroom, dressed in a robe, to my computer. These circumstances are all they were able to document. But despite their best efforts, there was no way for them to get a visual of my ideas and where they actually come from. The same video could have been made with the previous occupant of the house who shared none of my scientific ideas. He and I happened to use the same shower, eat in the same kitchen and sleep in the same bedroom, altogether sharing the same spaces (at different times) but with very different outcomes. Where do ideas come from?

Ideas originate from pregnant minds, just as babies come out from the bellies of their mothers. What makes a mind fertile? For one, it is the freedom to venture without the confines of traditional thinking or the burden of practical concerns. If a quantum system is probed too often, it tends to stay in the same state. The same is true for the mind of an individual if it is interrupted too often by others. Immersing oneself in the trivia of common wisdom resembles to reading yesterday's news in the daily newspaper, with no prospect for making a difference. Senior researchers aim to establish echo chambers in which their voices are heard loud and clear through their group members. This is an antidote to pregnancies with new ideas. Early career scientists might not fulfil their discovery potential if they accept the limits established by their mentors. Innovation occurs when researchers deviate from group thinking or fashion.

By its nature, persistent conservatism is ultimately doomed to a culture shock. In December 2013, I gave a pedagogical lecture on the topic of "Gravitational Wave Astrophysics" to students at the 30th Jerusalem winter school on Early Galaxy Formation. Within ten minutes into my lecture, a young lecturer at the school who specializes in traditional astronomy raised his hand and asked: "Why are you wasting the time of these students? We all know that this field will not be of use to them in their careers." In September 2015, while many of the same students were still working on their PhDs, LIGO discovered gravitational waves from the black holes merger GW150914. The subsequent detection of electromagnetic counterparts to the neutron stars merger GW170817 ushered a new era of multi-messenger astronomy, and the lecturer's prophecy was demonstrated to be officially wrong by the announcement of the 2017 Nobel prize in Physics. In hindsight, this blunder might not be surprising. Think about how riders of horse carriages viewed Ford's Model T car or how the executives of Britannica viewed Wikipedia in its early days.

Another "anti-pregnancy pill" involves the mundane burdens of life that consume the mind with practical concerns. This explains why I tend to get ideas in the shower, as the shower offers me vacation time from interruptions associated with my duties as chair of an Astronomy department and director of two centers. Without vacations from distractions, ideas are scarce. There are many alternatives to the shower that would offer the same fundamental benefit; vacations are sometimes defined by what they escape from more than by what they offer as a substitute.

How could we cultivate an environment that nourishes ideas? The recipe starts with creating a culture that encourages informal questioning and inquiry, tolerates mistakes and promotes innovation. For example, the Socratic method of dialogues[1] - which encouraged critical thinking and challenged authority, led to a rich literature of insights in philosophy and ethics, and is suitable also to science. But even fertile soil cannot guarantee blossoming vegetation without seeds. How could we seed an academic environment with ideas?

Ideas often originate from dialogues in which an individual hears about a challenge and recognizes a new path for solving it. It is therefore crucial to create a space in which challenges are discussed openly and without fear, stimulating new solutions. An excellent historic example was Bell Labs which for decades in the mid 20th century, assembled creative physicists and engineers into a single corridor, where their daily conversations led to the inventions of radio astronomy, the transistor, photovoltaic cells, the laser, the discovery of the cosmic microwave background, CCDs and many other breakthroughs[2]. It is essential to include young people in the conversation, since they lack baggage and are willing to march into uncharted territories.

The desire to create innovative environments extends well beyond academia, as it carries great financial benefits to businesses, such as Amazon, Google, Apple, Microsoft or Space X. Yet, the most beautiful discoveries in science occur for free and are not designed by corporate boards. It is well known that if you wish to obtain a traditional result with little variance, all you need to do is assemble a large committee; this outcome is guaranteed by the central limit theorem of statistics.

Finally, a word of caution for innovators. It is not enough to plant the seeds. Once born - ideas need to be attended to and developed, for the same reason that babies need to be fed in order to mature as adults. Many excellent ideas were lost due to neglect. In analogy with start-up companies, a good idea must be followed by a feasibility study which evaluates its promise and nurtures its further growth if it appears promising. Risks are inevitable, since in the dense fog of innovation you cannot tell if you are facing a plateau, a steep hill or a cliff just a few steps ahead. But once successful, a single excellent idea could be worth the investment in a hundred failed ones. And certainly worth the full attention of a Dutch TV crew.

## ABOUT THE AUTHOR

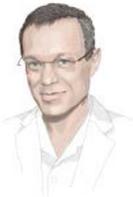

**Abraham Loeb**

Abraham Loeb is chair of the astronomy department at Harvard University, founding director of Harvard's Black Hole Initiative and director of the Institute for Theory and Computation at the Harvard-Smithsonian Center for Astrophysics. He chairs the Board on Physics and Astronomy of the National Academies and the advisory board for the Breakthrough Starshot project.

*Illustration credit:* Nick Higgins